\documentstyle[12pt]{article}

\font\twlgot =eufm10 scaled \magstep1
\font\egtgot =eufm8
\font\sevgot =eufm7

\font\twlmsb =msbm10 scaled \magstep1
\font\egtmsb =msbm8
\font\sevmsb =msbm7

\newfam\gotfam
\def\pgot{\fam\gotfam\twlgot}
\textfont\gotfam\twlgot
\scriptfont\gotfam\egtgot
\scriptscriptfont\gotfam\sevgot
\def\got{\protect\pgot}

\newfam\msbfam
\textfont\msbfam\twlmsb
\scriptfont\msbfam\egtmsb
\scriptscriptfont\msbfam\sevmsb

\def\pBbb{\relax\ifmmode\expandafter\Bb\else\typeout{You cann't use
Bbb in text mode}\fi}
\def\Bb #1{{\fam\msbfam\relax#1}}

\def\thebibliography#1{\bigskip\section*{\large
\bf References\\}\list
  {[\arabic{enumi}]}{\settowidth\labelwidth{#1}\leftmargin\labelwidth
    \advance\leftmargin\labelsep
    \usecounter{enumi}}
    \def\newblock{\hskip .11em plus .33em minus .07em}
    \sloppy\clubpenalty4000\widowpenalty4000
    \sfcode`\.=1000\relax}

\def\op#1{\mathop{{\it\fam0} #1}\limits}

\newcommand{\beq}{\begin{equation}}
\newcommand{\eeq}{\end{equation}}
\newcommand{\ben}{\begin{eqnarray}}
\newcommand{\een}{\end{eqnarray}}
\newcommand{\be}{\begin{eqnarray*}}
\newcommand{\ee}{\end{eqnarray*}}
\newcommand{\bea}{\begin{eqalph}}
\newcommand{\eea}{\end{eqalph}}

\newcommand{\cG}{{\got g}}

\newcommand{\gj}{{\got J}}

\newcommand{\gS}{{\got S}}

\newcommand{\gF}{{\got F}}

\newcommand{\gP}{{\got P}}

\newcommand{\cP}{{\cal P}}

\newcommand{\cL}{{\cal L}}

\newcommand{\cF}{{\cal F}}

\newcommand{\cK}{{\cal K}}

\newcommand{\cS}{{\cal S}}

\newcommand{\bL}{{\bf L}}

\newcommand{\al}{\alpha}
\newcommand{\bt}{\beta}
\newcommand{\dl}{\delta}
\newcommand{\la}{\lambda}

\newcommand{\vf}{\varphi}

\newcommand{\x}{\xi}
\newcommand{\om}{\omega}

\newcommand{\m}{\mu}

\newcommand{\g}{\gamma}

\newcommand{\e}{\epsilon}
\newcommand{\ve}{\varepsilon}
\newcommand{\th}{\theta}

\newcommand{\si}{\sigma}

\newcommand{\w}{\wedge}

\newcommand{\wt}{\widetilde}

\newcommand{\ol}{\overline}

\newcommand{\dr}{\partial}

\newcommand{\ot}{\otimes}
\newcommand{\ap}{\approx}

\newenvironment{eqalph}{\stepcounter{equation}
\setcounter{equationa}{\value{equation}}
\setcounter{equation}{0}

\begin{eqnarray}}{\end{eqnarray}\setcounter{equation}{\value{equationa}}}

\newcounter{example}
\newcounter{remark}
\newcounter{theorem}
\newcounter{proposition}
\newcounter{lemma}
\newcounter{corollary}
\newcounter{definition}

\def\thedefinition{\arabic{definition}}

\newcommand{\mar}[1]{}

\begin{document}
\hbox{}

{\parindent=0pt

{\large \bf Energy-momentum conservation laws in higher-dimensional 
Chern--Simons models} 
\bigskip 

{\bf G.Sardanashvily}

\medskip

\begin{small}

Department of Theoretical Physics, Moscow State University, 117234
Moscow, Russia

E-mail: sard@grav.phys.msu.su

URL: http://webcenter.ru/$\sim$sardan/
\bigskip

{\bf Abstract.}
Though a Chern--Simons $(2k-1)$-form is not gauge-invariant and it
depends on a background connection, this form seen as a 
Lagrangian of gauge theory on 
a $(2k-1)$-dimensional manifold leads to the energy-momentum conservation 
law. 
\end{small}
}
\bigskip
\bigskip

The local Chern--Simons (henceforth CS) form seen as a Lagrangian of
the gauge theory on a 3-dimensional manifold is well known to lead
to the local conservation law of the canonical energy-momentum tensor.
Generalizing this result, we show that a global higher-dimensional 
CS gauge theory admits
an energy-momentum conservation law in spite of the fact that its
Lagrangian depends on a background gauge potential and that one can not
ignore its gauge non-invariance. The CS
gravitation theory here is not considered (e.g., \cite{al,bor2}).

We derive Lagrangian energy-momentum
conservation laws from the first variational formula (e.g.,
\cite{book,book00,sard97,epr,epr1}). Let us consider a first order
field theory on a fibre bundle $Y\to X$ over an $n$-dimensional base $X$.
Its Lagrangian is defined as a density $L=\cL d^nx$ on the first order
jet manifold $J^1Y$ of sections of $Y\to X$. 
Given bundle coordinates $(x^\la,y^i)$ on a fibre bundle $Y\to
X$, its first and second order jet manifolds $J^1Y$ and 
$J^2Y$ are equipped with the
adapted coordinates $(x^\la,y^i,y^i_\m)$ and $(x^\la,y^i,y^i_\m,
y^i_{\la\m})$, respectively. 
We will use the notation $\om=d^nx$ and
$\om_\la=\dr_\la\rfloor \om$.

Given a Lagrangian 
$L$
on $J^1Y$, the corresponding Euler--Lagrange operator reads
\mar{305}\beq
\dl L= \dl_i\cL \th^i\w\om=(\dr_i\cL- d_\la\dr^\la_i)\cL \th^i\w\om,
\label{305} 
\eeq
where $\th^i=dy^i- y^i_\la dx^\la$ are contact forms and $d_\la=\dr_\la
+y^i_\la\dr_i +y^i_{\la\m}\dr_i^\m$ are the total 
derivatives, which yield the total differential 
\mar{r73}\beq
d_H\vf=dx^\la\w d_\la\vf \label{r73}
\eeq
acting on the pull-back of exterior forms on $J^1Y$ onto $J^2Y$.
The kernel
Ker$\,\dl L\subset J^2Y$ of the Euler--Lagrange operator (\ref{305})
defines the Euler--Lagrange 
equations
$\dl_i\cL=0$. A Lagrangian $L$ is said to be variationally trivial if 
$\dl L=0$. This property holds iff
$L=h_0(\vf)$, 
where
$\vf$ is a closed $n$-form on $Y$ and $h_0$ is the horizontal projection 
\mar{r71}\beq
h_0(dx^\la)=dx^\la, \qquad h_0(dy^i)=y^i_\la dx^\la, \qquad h_0(dy^i_\m)=
y_{\la\m}^idx^\la. \label{r71}
\eeq
The relation $d_H\circ h_0=h_0\circ d$ holds.

Any projectable vector field 
\mar{e1}\beq
u=u^\la(x^\m)\dr_\la +u^i(x^\m,y^j)\dr_i \label{e1}
\eeq
on $Y\to X$ is the infinitesimal
generator of a local one-parameter group of bundle automorphisms of
$Y\to X$, and {\it vice versa}. Its prolongation onto $J^1Y$ is
\mar{1.21}\beq
J^1u=u^\la\dr_\la + u^i\dr_i +(d_\la u^i-y^i_\m\dr_\la u^\m)\dr^\la_i.
\label{1.21} 
\eeq
Then, the Lie derivative of a Lagrangian $L$ along $J^1u$ reads
\mar{r8}\beq
\bL_{J^1u}L=J^1u\rfloor dL +d(J^1u\rfloor L)=[\dr_\la(u^\la \cL) +
u^i\dr_i\cL +(d_\la u^i-y^i_\m\dr_\la u^\m)\dr^\la_i\cL]\om. \label{r8}
\eeq
The first variational
formula provides its canonical decomposition
\mar{bC30'}\ben
&& \bL_{J^1u}L= 
 u_V\rfloor \dl L + d_Hh_0(u\rfloor H_L) \label{bC30'} =\\
&& \qquad   (u^i-y^i_\m u^\m )\dl_i\cL\om - 
d_\la[(u^\m y^i_\m -u^i)\dr^\la_i\cL -u^\la\cL]\om, \nonumber
\een
where $u_V=(u\rfloor\th^i)\dr_i$, $H_L$
is the Poincar\'e--Cartan form, and 
\mar{Q30}\beq
\gj_u =-h_0(u\rfloor H_L)=\gj_u^\la\om_\la
=[(u^\m y^i_\m-u^i)\dr^\la_i\cL -u^\la\cL]\om_\la 
\label{Q30}
\eeq
is the symmetry current along $u$. On Ker$\,\dl L$, 
the first variational formula (\ref{bC30'}) leads to the weak 
equality
\mar{J4}\ben
&& \bL_{J^1u}L\ap-d_H\gj_u, \label{J4}\\
&& \dr_\la(u^\la\cL)+
u^i\dr_i\cL +(d_\la u^i -y^i_\m\dr_\la u^\m)\dr^\la_i\cL \ap 
- d_\la[\c(u^\m y^i_\m -u^i)\dr^\la_i\cL -u^\la\cL]. \nonumber
\een
Let the Lie derivative (\ref{r8}) reduces to
the total differential 
\mar{r4}\beq
\bL_{J^ru}L=d_H\si, \label{r4}
\eeq
e.g., if $L$ is a variationally trivial Lagrangian or
$L$ is invariant under a
one-parameter group
of bundle automorphisms of $Y\to X$ generated by $u$.
Then, the weak equality (\ref{J4}) takes the form 
\mar{r5}\beq
0\ap -d_H(\gj_u+\si), \label{r5}
\eeq
regarded as a conservation law of the modified symmetry current
$\ol\gj=\gj_u+\si$. 

It is readily observed that symmetry currents (\ref{Q30}) obey the relation
\mar{e4}\beq
\gj_{u+u'}=\gj_u +\gj_{u'}. \label{e4}
\eeq
Note that any projectable vector field $u$
(\ref{e1}), projected onto 
a vector field $\tau=u^\la\dr_\la$ on $X$, can be written as the sum
\mar{e3}\beq
u=\wt\tau +(u-\wt\tau) \label{e3}
\eeq
of some lift $\wt\tau=u^\la\dr_\la +\wt\tau^i\dr_i$
of $\tau$ onto $Y$ and the vertical vector field 
$u-\wt\tau$ on $Y$. 
The symmetry current $\gj_u$ (\ref{Q30}) along a vertical
vector field $u=u^i\dr_i$ is the Noether current
$\gj^\la_u =-u^i\dr^\la_i\cL$.
The current $\gj_{\wt\tau}$ (\ref{Q30}) along a lift $\wt\tau$ onto $Y$
of a vector field $\tau=\tau^\la\dr_\la$ on $X$ is the
energy-momentum current \cite{fer,book,got,sard97}. 
Then the decompositions (\ref{e4}) and (\ref{e3}) 
show that any current
$\gj_u$ (\ref{Q30}) along a projectable vector field $u$ on a fibre
bundle $Y\to X$ can be represented by a sum of an energy-momentum
current and a Noether one. 
Of course, different lifts $\wt\tau$ and $\wt\tau'$ onto $Y$ of a vector field
$\tau$ on $X$ lead to different energy-momentum currents $\gj_{\wt\tau}$
and $\gj_{\wt\tau'}$, whose difference $\gj_{\wt\tau}-\gj_{\wt\tau'}$
is the Noether current along the vertical vector field
$\wt\tau-\wt\tau'$ on $Y$. 

Note that there exists the category of so called natural fibre bundles
$T\to X$ (e.g., tensor bundles) which admit the canonical lift $\wt\tau$ 
of vector fields $\tau$ on $X$. Such a lift is the infinitesimal generator of a
one-parameter group of general covariant transformations of $T$. This
is the case of gravitation theory which we here do not concern.

Let us turn to gauge theory of principal connections on a
principal bundle $P\to X$ with a structure  
Lie group $G$. Let $J^1P$ be the first order jet manifold of $P\to X$
and 
\mar{r43}\beq
C=J^1P/G\to X \label{r43}
\eeq
the quotient of $P$ with respect to the canonical action of $G$ on
$P$ (e.g., \cite{book,book00,epr}).  
There is one-to-one correspondence between the principal connections on
$P\to X$ and the sections of the fibre bundle $C$
(\ref{r43}), called the 
connection bundle. Given an atlas $\Psi$ of $P$, the connection bundle
$C$ is provided with bundle 
coordinates $(x^\la, a^r_\m)$ such that, for any its section $A$, the
local functions $A^r_\m=a^r_\m\circ A$ are coefficients of the
familiar local connection form. From the physical viewpoint, $A$
is a gauge potential.

The infinitesimal 
generators of local one-parameter groups of 
automorphism of a principal bundle $P\to X$
are $G$-invariant projectable vector fields on $P\to X$. There
is one-to-one correspondence between these vector fields and the sections of
the quotient 
\mar{r44}\beq
T_GP=TP/G\to X \label{r44}
\eeq
of the tangent bundle $TP$ of $P$ with respect to the
canonical action of $G$ on $P$.
Given a basis $\{\e_r\}$ for the
right Lie algebra $\cG_r$ of the group $G$, let $\{\dr_\la, e_r\}$ 
be the corresponding
fibre bases for the vector bundle $T_GP$. Then, a section $\x$
of $T_GP\to X$ reads
\mar{e10}\beq
\x =\x^\la\dr_\la + \x^r e_r. \label{e10}
\eeq
The infinitesimal generator
of a one-parameter group  of vertical automorphisms is a
$G$-invariant vertical vector field on $P$ identified
to a section $\x=\x^r e_r$ of the quotient 
\be
V_GP=VP/G\subset T_GP 
\ee
of the vertical tangent
bundle $VP$ of $P\to X$ by the canonical action of $G$ on $P$.
The Lie bracket of two sections
$\x$ and $\eta$ 
of the vector bundle $T_GP\to X$ reads
\be
  [\x,\eta ]=(\x^\m\dr_\m\eta^\la - \eta^\m\dr_\m\x^\la)\dr_\la
  +(\x^\la\dr_\la\eta^r - \eta^\la\dr_\la\x^r +
c_{pq}^r\x^p\eta^q) e_r, 
\ee
where $c_{pq}^r$ are the structure constants of the 
Lie algebra ${\got g}_r$. Putting $\xi^\la=0$ and $\eta^\m=0$, we
obtain the Lie bracket 
\mar{1129'}\beq
[\x,\eta]= c_{pq}^r\x^p\eta^q e_r \label{1129'}
\eeq
of sections of the vector bundle $V_GP\to X$.
A glance at the expression (\ref{1129'}) shows
that the typical fibre of $V_GP\to X$ is the 
Lie algebra ${\got g}_r$. The structure group $G$ acts on ${\got g}_r$ by the 
adjoint representation. 

Note that the connection bundle $C$ (\ref{r43}) is an affine bundle
modelled over the vector bundle $T^*X\ot V_GP$, and elements of $C$ are
represented by local $V_GP$-valued 1-forms $a^r_\m
dx^\m\ot e_r$. 
Bundle automorphisms of a principal bundle $P\to X$ generated by the
vector field 
(\ref{e10}) induce bundle automorphisms of the connection bundle $C$ whose 
generator is 
\mar{e11}\beq
\x_C=\x^\la\dr_\la+ (\dr_\m\x^r +
c^r_{pq}a^p_\m\x^q-a^r_\nu\dr_\m\x^\nu)\dr^\m_r. \label{e11}
\eeq

The connection bundle $C\to X$ admits the canonical $V_GP$-valued 2-form
\mar{r60}\beq
\gF=(da^r_\m\w dx^\m +\frac12 c^r_{pq}a^p_\la a^q_\m dx^\la\w dx^\m)\ot
e_r, \label{r60}
\eeq
which is the curvature of the canonical connection on the principal
bundle $C\times P\to C$ (e.g., \cite{book00}). 
Given a section $A$ of $C\to X$, the pull-back  
\mar{r47'}\beq
F_A=A^*\gF=\frac12 F^r_{\la\m}dx^\la\w dx^\m\ot e_r,
\qquad  F^r_{\la\m}=\dr_\la A^r_\m-\dr_\m A^r_\la +c^r_{pq}A^p_\la A^q_\m,
\label{r47'}
\eeq
of $\gF$ onto $X$ is the strength form of a gauge potential $A$.  

Let $I_k(\e)=b_{r_1\ldots r_k}\e^{r_1}\cdots \e^{r_k}$ be a $G$-invariant
polynomial of degree $k>1$ on the Lie algebra $\cG_r$ written with
respect to its basis $\{\e_r\}$, i.e.,
\be
I_k(\e)=\op\sum_j b_{r_1\ldots r_k}\e^{r_1}\cdots c^{r_j}_{pq}e^p\cdots
\e^{r_k}=
kb_{r_1\ldots r_k}c^{r_1}_{pq}\e^p\e^{r_2}\cdots \e^{r_k}=0.
\ee
Let us associate to $I(\e)$ the $2k$-form 
\mar{r61}\beq
P_{2k}(\gF)=b_{r_1\ldots r_k}\gF^{r_1}\w\cdots\w \gF^{r_k} \label{r61}
\eeq
on $C$. It is a closed form which is invariant under vertical
automorphisms of $C$.
Let $A$ be a section of $C\to X$. Then, the pull-back 
\mar{r63}\beq
P_{2k}(F_A)=A^*P_{2k}(\gF) \label{r63}
\eeq
of $P_{2k}(\gF)$ is a closed characteristic form on $X$. Recall that
the de Rham cohomology of $C$ equals that of $X$ since $C\to X$
is an affine bundle. It follows that $P_{2k}(\gF)$ and $P_{2k}(F_A)$ possess
the same  
cohomology class 
\mar{r62}\beq
[P_{2k}(\gF)]=[P_{2k}(F_A)] \label{r62}
\eeq
for any principal connection $A$. Thus, $I_k(\e)\mapsto
[P_{2k}(F_A)]\in H^*(X)$ 
is the familiar Weil homomorphism.

Let $B$ be a fixed section of the connection bundle $C\to X$.
Given the characteristic form $P_{2k}(F_B)$ (\ref{r63}) on $X$, let the same
symbol stand for its pull-back onto $C$. By virtue of the equality 
(\ref{r62}), the difference $P_{2k}(\gF)-P_{2k}(F_B)$ is an exact form on $C$.
Moreover, similarly to the well-known transgression formula on a
principal bundle $P$, one can obtain the following transgression formula
on $C$:
\mar{r64,5}\ben
&& P_{2k}(\gF)-P_{2k}(F_B)=d\gS_{2k-1}(B), \label{r64}\\
&&  \gS_{2k-1}(B)=k\op\int^1_0 \gP_{2k}(t,B)dt, \label{r65}\\
&& \gP_{2k}(t,B)=b_{r_1\ldots r_k}(a^{r_1}_{\m_1}-B^{r_1}_{\m_1})dx^{\m_1}\w
\gF^{r_2}(t,B)\w\cdots \w \gF^{r_k}(t,B),\nonumber\\
&& \gF^{r_j}(t,B)=
[d(ta^{r_j}_{\m_j} +(1-t)B^{r_j}_{\m_j})\w dx^{\m_j} +\nonumber\\
&& \qquad \frac12c^{r_j}_{pq}
(ta^p_{\la_j} +(1-t)B^p_{\la_j})(ta^q_{\m_j} +(1-t)B^q_{\m_j})dx^{\la_j}\w 
dx^{\m_j}]\ot e_r.
\nonumber
\een
Its pull-back by means of a section $A$ of $C\to X$ gives
the transgression formula
\be
P_{2k}(F_A)-P_{2k}(F_B)=d S_{2k-1}(A,B)
\ee
on $X$. For instance, if $P_{2k}(F_A)$ is the characteristic Chern $2k$-form,
then $S_{2k-1}(A,B)$ is the familiar CS $(2k-1)$-form. 
Therefore, we agree to call $\gS_{2k-1}(B)$
(\ref{r65}) the CS form on the connection bundle $C$. 
In particular, one can choose the local section $B=0$. 
Then, $\gS_{2k-1}=\gS_{2k-1}(0)$ is
the local CS form. Let $S_{2k-1}(A)$ denote its pull-back 
onto $X$ by means of a section $A$ of $C\to X$. Then, the CS form
$\gS_{2k-1}(B)$ admits the decomposition
\mar{r75}\beq
\gS_{2k-1}(B)=\gS_{2k-1} -S_{2k-1}(B) +dK_{2k-1}(B). \label{r75}
\eeq

Let $J^1C$ be the first order jet manifold of the
connection bundle $C\to X$ equipped with the adapted coordinates
$(x^\la, a^r_\m, a^r_{\la\m})$. 
Let us consider the pull-back of the CS form (\ref{r65}) onto $J^1C$
denoted by the same symbol $\gS_{2k-1}(B)$, and let 
\mar{r74}\beq
\cS_{2k-1}(B)=h_0 \gS_{2k-1}(B) \label{r74}
\eeq
be its horizontal projection.
This is given by the formula
\be
&& \cS_{2k-1}(B)=k\op\int^1_0 \cP_{2k}(t,B)dt, \\
&& \cP_{2k}(t,B)=b_{r_1\ldots r_k}(a^{r_1}_{\m_1}-B^{r_1}_{\m_1})dx^{\m_1}\w
\cF^{r_2}(t,B)\w\cdots \w \cF^{r_k}(t,B),\\
&& \gF^{r_j}(t,B)= \frac12[
ta^{r_j}_{\la_j\m_j} +(1-t)\dr_{\la_j}B^{r_j}_{\m_j}
- ta^{r_j}_{\m_j\la_j} -(1-t)\dr_{\m_j}B^{r_j}_{\la_j}+\\
&& \qquad \frac12c^{r_j}_{pq}
(ta^p_{\la_j} +(1-t)B^p_{\la_j})(ta^q_{\m_j} +(1-t)B^q_{\m_j}]dx^{\la_j}\w 
dx^{\m_j}\ot e_r.
\ee
The decomposition (\ref{r75}) 
induces the decomposition 
\mar{r100}\beq
\cS_{2k-1}(B)=\cS_{2k-1} -S_{2k-1}(B) +d_H\cK_{2k-1}(B), 
\qquad \cK_{2k-1}(B)=h_0 K_{2k-1}(B). \label{r100}
\eeq

Now, let us consider the CS gauge model on a $(2k-1)$-dimensional base
manifold $X$ whose Lagrangian 
\mar{r80}\beq
L_{\rm CS}=\cS_{2k-1}(B) \label{r80}
\eeq
is the CS form (\ref{r74}) on $J^1C$. Let 
\mar{r47}\beq
\x_C=(\dr_\m\x^r +
c^r_{pq}a^p_\m\x^q)\dr^\m_r \label{r47}
\eeq
be a vertical vector field on $C$. We have shown that 
\mar{r102}\beq
\bL_{J^1\xi_C} \cS_{2k-1}(B)=d_H \si \label{r102}
\eeq
\cite{epr2}. 
As a consequence, the CS gauge model with the Lagrangian (\ref{r80}) admits
the Noether conservation law 
\be
0\ap -d_H(\gj_\xi +\si),
\ee
where
\be
\gj_\xi=\dr_r^{\la\m}\cS_{2k-1}(B) (\dr_\m\x^r +
c^r_{pq}a^p_\m\x^q)
\ee
is the symmetry current along the vertical vector field $\xi_C$ (\ref{r47}).

Using this fact, let us study energy-momentum conservation laws in the CS
gauge model with the Lagrangian (\ref{r80}). Let $B'$ be some section
of the connection bundle $C\to X$. 
Given a vector field $\tau$ on $X$, there
exists its lift
\mar{e25}\beq
\wt\tau_{B'}=\tau^\la\dr_\la +[\dr_\m(\tau^\nu B'^r_\nu)+c^r_{pq}
a^p_\m (\tau^\nu B'^q_\nu) - a^r_\nu\dr_\m \tau^\nu]\dr^\m_r. \label{e25}
\eeq
onto the connection bundle $C\to X$ 
\cite{book,book00,sard97}. 
Comparing the expressions (\ref{e11}) and (\ref{e25}), one easily
observes that the lift $\wt\tau_{B'}$ is the infinitesimal generator of 
bundle automorphisms of $C$ with parameters $\x^\la=\tau^\la$,
$\x^r=\tau^\nu B^r_\nu$.
Let us bring locally the vector field (\ref{e25}) into the sum
\mar{mos54}\beq
\wt\tau_{B'} =\wt\tau_1+\wt\tau_2= [\tau^\la\dr_\la 
- a^r_\nu\dr_\m \tau^\nu\dr^\m_r] + 
[\dr_\m(\tau^\nu B'^r_\nu)+c^r_{pq}
a^p_\m (\tau^\nu B'^q_\nu)]\dr^\m_r. \label{mos54}
\eeq
One can think of $\wt\tau_1$ as being the local infinitesimal generator of
general covariant transformations of $C$, while $\wt\tau_2$ is a local
vertical vector field on $C\to X$ (cf. (\ref{r47})). Then, the weak
equality (\ref{J4}) for the Lie derivative
$\bL_{J^1\wt\tau}\cS_{2k-1}(B)$ reads
\mar{r103}\beq
\bL_{J^1\wt\tau_1}\cS_{2k-1}(B) + \bL_{J^1\wt\tau_2}\cS_{2k-1}(B)\ap
-d_H\gj_{\wt\tau}, \label{r103}
\eeq
where $\gj_{\wt\tau}$ is the energy-momentum current along $\wt\tau_{B'}$.
The second term in the left-hand side of this equality takes the form
(\ref{r102}), and the first one does so as follows. Using the
decomposition (\ref{r100}), one can write
\be
\bL_{J^1\wt\tau_1}\cS_{2k-1}(B)=\bL_{J^1\wt\tau_1}\cS_{2k-1}
+\bL_{J^1\wt\tau_1}(S_{2k-1}(B)+ d_H\cK_{2k-1}(B)).
\ee 
The first term in the right-hand side of this relation vanishes because
the local CS form $\cS_{2k-1}$ is invariant under general covariant
transformations. The second one takes the form (\ref{r4}) because 
$S_{2k-1}(B)+ d_H\cK_{2k-1}(B)$ is a variationally trivial Lagrangian.
Consequently, the weak equality (\ref{r103}) comes to the conservation law
\be
0\ap -d_H(\gj_{\wt\tau} +\si')
\ee
of the modified energy-momentum current.

For instance, let $G$ be a semi-simple group and $a^G$ 
the Killing form on $\cG_r$. Let 
\mar{r48}\beq
P(\gF)=\frac{h}{2}a^G_{mn}\gF^m\w \gF^n \label{r48}
\eeq
be the second Chern form up to a constant multiple. Given a section $B$ of
$C\to X$, the transgression formula
(\ref{r64}) on $C$ reads
\mar{r49}\beq
P(\gF)-P(F_B)=d\gS_3(B), \label{r49}
\eeq 
where $\gS_3(B)$ is the CS 3-form up to a constant multiple. 
Let us a consider the gauge model on a 3-dimensional base manifold
whose Lagrangian is 
\mar{r50}\ben
&&L_{\rm CS}=h_0(\gS_3(B))= \cS_3 -S_3(B) +d_H\cK_3(B)=
[\frac12 ha^G_{mn} \ve^{\al\bt\g}a^m_\al(\cF^n_{\bt\g} -\frac13
c^n_{pq}a^p_\bt a^q_\g)  \label{r50}\\
&& \qquad -\frac12 ha^G_{mn} \ve^{\al\bt\g}B^m_\al(F(B)^n_{\bt\g} -\frac13
c^n_{pq}B^p_\bt B^q_\g)
-d_\al(ha^G_{mn} \ve^{\al\bt\g}a^m_\bt B^n_\g)]d^3x, \nonumber\\
&& \cF=h_0\gF=\frac12 \cF^r_{\la\m}dx^\la\w dx^\m\ot e_r, \qquad
\cF^r_{\la\m}=a^r_{\la\m}-a^r_{\m\la} +c^r_{pq}a^p_\la a^q_\m.
\nonumber
\een
Since $(-S_3(B) +d_H\cK_3(B))$ is a variationally trivial local
Lagrangian, the first variational formula for the CS Lagrangian (\ref{r50})
reduces to that for the local CS Lagrangian 
\mar{r105}\beq
\cS_3=\frac12 ha^G_{mn} \ve^{\al\bt\g}a^m_\al(\cF^n_{\bt\g} -\frac13
c^n_{pq}a^p_\bt a^q_\g)d^3x. \label{r105}
\eeq
On-shell, it reads
\mar{r86}\beq
\bL_{J^1\wt\tau_2}\cS_3\ap - d_H\gj_{\wt\tau}, \label{r86}
\eeq
where $\gj_{\wt\tau}$ is the symmetry current (\ref{Q30}) of $\cS_3$
(\ref{r105}) along the vector field $\wt\tau_{B'}$ (\ref{mos54}).
A simple calculation gives 
\be
&& \bL_{J^1\wt\tau_2}\cS_3
=-d_\al[ha^G_{mn}\ve^{\al\bt\g}\dr_\bt(\tau^\nu B'^m_\nu) a^n_\g]d^3x,\\
&& \gj_{\wt\tau}^\al=-ha^G_{mn}\ve^{\al\bt\g}[a^m_\bt d_\g(\tau^\nu a^n_\nu)
+\dr_\bt(\tau^nu B'^m_\nu
+c^m_{pq}a^p_\bt\tau^\nu B'^q_\nu)a^n_\g]-\tau^\al\cS_3. 
\ee
Thus, we come to the conservation law 
of the modified energy-momentum current
\be
\ol\gj=\gj_{\wt\tau}^\al -
ha^G_{mn}\ve^{\al\bt\g}\dr_\bt(\tau^\nu B'^m_\nu) a^n_\g.
\ee


\begin{thebibliography}{ederf}


\bibitem{al} G.Allemandi, M.Francaviglia and M.Raiteri, Covariant
charges in Chern--Simons $AdS_3$
gravity, {\it Class. Quant. Grav.} {\bf 20} (2003) 483; {\it E-print
arXiv:} gr-qc/0211098. 

\bibitem{bor98} A.Borowiec, M.Ferraris and M.Francaviglia, Lagrangian
symmetries of Chern--Simons theories, {\it J. Phys. A} {\bf 31} (1998)
8823: {\it E-print arXiv:} hep-th/9801126.

\bibitem{bor2} A.Borowiec, M.Ferraris and M.Francaviglia, A covariant
formalism for Chern--Simons gravity, {\it J. Phys. A} {\bf 36} (2003) 2589;
{\it E-print arXiv}: math-ph/0301146.

\bibitem{fer} M.Ferraris and M.Francaviglia, Energy-momentum tensors and
stress tensors in geometric field theories, {\it J. Math.Phys.} 
{\bf 26} (1985) 1243.

\bibitem{book} G.Giachetta, L.Mangiarotti and G.Sardanashvily, {\it
New Lagrangian and Hamiltonian Methods in Field Theory} (World Scientific,
Singapore, 1997).

\bibitem{got} M.Gotay and J.Marsden, Stress-energy-momentum tensors and
the Belinfante--Rosenfeld formula, {\it Contemp. Mathem.} {\bf 132}
(1992) 367.

\bibitem{book00} L.Mangiarotti and G.Sardanashvily, {\it
Connections in Classical and Quantum Field Theory} (World Scientific,
Singapore, 2000).

\bibitem{sard97} G.Sardanashvily, Stress-energy-momentum tensor in
constraint field theories, {\it J. Math. Phys.} {\bf 38} (1997) 847.

\bibitem{epr} G.Sardanashvily, Ten lectures on jet manifold in
classical and quantum field theory, {\it E-print arXiv}: math-ph/0203040.

\bibitem{epr1} G.Sardanashvily, Energy-momentum conservation laws in gauge
theory with broken gauge invariance, {\it E-print arXiv}: hep-th/0203275.

\bibitem{epr2} G.Sardanashvily, Gauge conservation laws in
higher-dimensional Chern--Simons models, {\it E-print arXiv}: hep-th/0303059.

\end{thebibliography}
\end{document}